\documentclass[12pt]{iopart}

\usepackage{epsfig}
\usepackage{amssymb}

\newtheorem{thm}{Theorem}

\begin{document}

\paper{Log-periodic drift oscillations in self-similar billiards}
\author{Felipe Barra}
\address{Deptartamento de F\'{\i}sica, Facultad de Ciencias
 F\'{\i}sicas y Matem\'aticas, Universidad de Chile, Casilla 487-3,
 Santiago Chile}
\author{Nikolai Chernov}
\address{Department of Mathematics, University of Alabama at
Birmingham, Birmingham, AL 35294}
\author{Thomas Gilbert}
\address{Center for Nonlinear Phenomena and Complex Systems,
 Universit\'e Libre  de Bruxelles, CP~231, Campus Plaine, B-1050
 Brussels, Belgium}
\date{\today}

\begin{abstract}
We study a particle moving at unit speed in a self-similar Lorentz
billiard channel; the latter consists of an infinite sequence of
cells which are identical in shape but growing exponentially in size,
from left to right. We present numerical computation of the drift
term in this system and establish the logarithmic periodicity of
the corrections to the average drift.
\end{abstract}

\submitto{\NL}

\pacs{05.45.-a,05.70.Ln,05.60.-k}

\ead{fbarra@dfi.uchile.cl, chernov@math.uab.edu, thomas.gilbert@ulb.ac.be}

\maketitle

\section{Introduction}

Through the last decades, the theory of hyperbolic dynamical systems
has become a cornerstone in the foundation of non-equilibrium
statistical mechanics. The interest in hyperbolic models stems from
relations between macroscopic features characteristic of
irreversible phenomena, such as entropy production or transport
coefficients, and dynamical properties, such as Lyapunov exponents
\cite{H86,EM90,Gas98,Dor99}.

Billiard models, whose dynamics are conveniently described by the
respective collision map, have proved extremely useful in this
regard \cite{CM06}. The simplest example is the two-dimensional
periodic Lorentz gas with finite horizon, which exhibits diffusion
without a drift \cite{BSC91}. In a statistically stationary state,
this model represents a mechanical system at equilibrium, whose
distribution, measured along the boundary of the scattering discs,
is uniform in the position and normal velocity angles. Meanwhile the
process of relaxation to equilibrium itself is characterized by
deterministic hydrodynamic modes of diffusion with fractal
properties \cite{GCGD01}. One can also induce a non-equilibrium
stationary state on a cylindrical version of this billiard by
coupling it to stochastic reservoirs of point-particles at its
boundaries \cite{Gas97}; these are the so-called flux boundary
conditions. In this case, a difference between the chemical
potentials of the reservoirs will transform to a linear gradient of
density across the system, which is responsible for a steady
current, given according to Fick's law of diffusion, and positivity
of the entropy production, see \cite{Gas98}.

Whereas the previous examples of billiards are Hamiltonian systems
which preserve the phase volume, one can also consider dissipative
periodic billiards driven out of equilibrium by the action of an
external field. Thus the Gaussian iso-kinetic periodic Lorentz gas
is similar to the usual periodic Lorentz gas, but for the action of
a thermostated uniform external field which bends the trajectories
in the direction of the field while keeping the particle's kinetic
energy constant \cite{H86}. This field therefore induces a
non-equilibrium stationary state to which is associated a constant
drift current, according to Ohm's law \cite{CELS93}. Moreover, due
to the dissipation induced by the thermostat, phase-space volume is,
on average, contracted. Therefore the sum of the Lyapunov exponents
is strictly negative in the presence of the field and one can in
fact identify this sum as minus the entropy production rate
\cite{R96}.

Gaussian iso-kinetic dynamics are different from Hamiltonian
dynamics in that the latter preserve the total energy, whereas the
former preserve the kinetic energy. One can nevertheless describe
the Gaussian iso-kinetic dynamics by a Hamiltonian formalism, as
shown in \cite{DM96}. A general theorem due to Wojtkowski \cite{W00}
stipulates that Gaussian iso-kinetic trajectories are geodesic lines
of the so-called torsion free connection (also known as the Weyl
connection). The Gaussian iso-kinetic Lorentz gas can thus be
conformally transformed into a distorted billiard table on which
trajectories become straight lines; besides, the conformal
transformation preserves the specular character of the collision
laws \cite{BG07b}. Although the trajectories thus transformed do not
have constant speed anymore, one can introduce, for any given
trajectory, a time-reparametrization under which the speed does
remain constant.

A volume-preserving billiard in a distorted channel is therefore a
natural generalization of iso-kinetic dynamics in a field-driven
Lorentz channel. Ignoring the strict periodicity of the latter
billiard, one may put aside the problem of time scales altogether
and consider the dynamics of independent point-particles moving at
constant speed in the new geometry. Such self-similar billiard
channels were introduced in \cite{BGR06,BG07a}; they consist of an
infinite sequence of (nonidentical) two-dimensional cells that are
attached together and make a (nonuniform) one-dimensional channel.
The cells are identical in shape, but their sizes are scaled by a
common factor. As a particle moves from one cell to a neighbouring
one, its velocity remains unchanged while the length scales are
expanded (or contracted), so that the time scales between collisions
change accordingly. A noticeable property of these billiards concern
their long-term statistics; even though their dynamics preserve
phase-space volumes, their statistics are characterized by a
non-equilibrium stationary state with fractal properties and a
drift, quite similar to the Gaussian iso-kinetic Lorentz gas itself.
Close to equilibrium (when the scaling between neighbouring cells is
close to unity), one can relate the constant drift velocity to the
diffusion coefficient of the equilibrium system, which is none but
the usual Lorentz channel, see \cite{BG07a}.

In a recent paper \cite{CD07}, Chernov and Dolgopyat unveiled a
remarkable feature of the drift term of self-similar billiards.
Namely that the linear growth of the particle's average displacement
with respect to time is periodic on logarithmic time scales, or
log-periodic, with a period specified by the scaling parameter. This
term may therefore display periodic oscillations. The purpose of
this paper is to provide further insight into this peculiar
phenomenon, which actually stems from the discrete scale invariance
of self-similar billiards \cite{Sor98}. We will offer numerical
evidence which demonstrates the existence of log-periodic drift
oscillations for self-similar billiards which are sufficiently far
away from equilibrium. Closer to equilibrium, these oscillations
are, at least numerically, found to vanish. Thus, for all practical
purposes, we may assume the drift is constant close to equilibrium,
and grows as a linear function of the scaling parameter. Further
away from equilibrium, this linear dependence of the average drift
in the scaling exponent remains valid, despite the presence of drift
oscillations.

The paper is organized as follows. The model is described in section
\ref{sec.model}, which is a slight modification of the models
previously studied in \cite{BGR06,BG07a}. Statistical properties of
the model are briefly discussed in section \ref{sec.erg}. The
characteristics of stationary drift, with numerical results, are
presented in section \ref{sec.osc}. Conclusions and perspectives are
offered in section \ref{sec.con}.

\section{Description of the model\label{sec.model}}

In order to display the log-periodicity of the drift function of self-similar
billiards discussed in section \ref{sec.osc}, which was predicted in
\cite{CD07}, we will modify the self-similar billiard that was defined
earlier in \cite{BGR06,BG07a}, and use a unit
cell rather similar to the one obtained after conformal transformation of
the iso-kinetic Lorentz channel \cite{BG07b}. The resulting cell has
enhanced symmetry in that all the discs are now equivalent. The main
motivation for introducing this modification is that it allows to take much
greater values of the scaling parameter within the constraints imposed to
guarantee ergodicity.

\begin{figure}[htb]
\begin{center}
\includegraphics[angle=0,width=.25\textwidth]{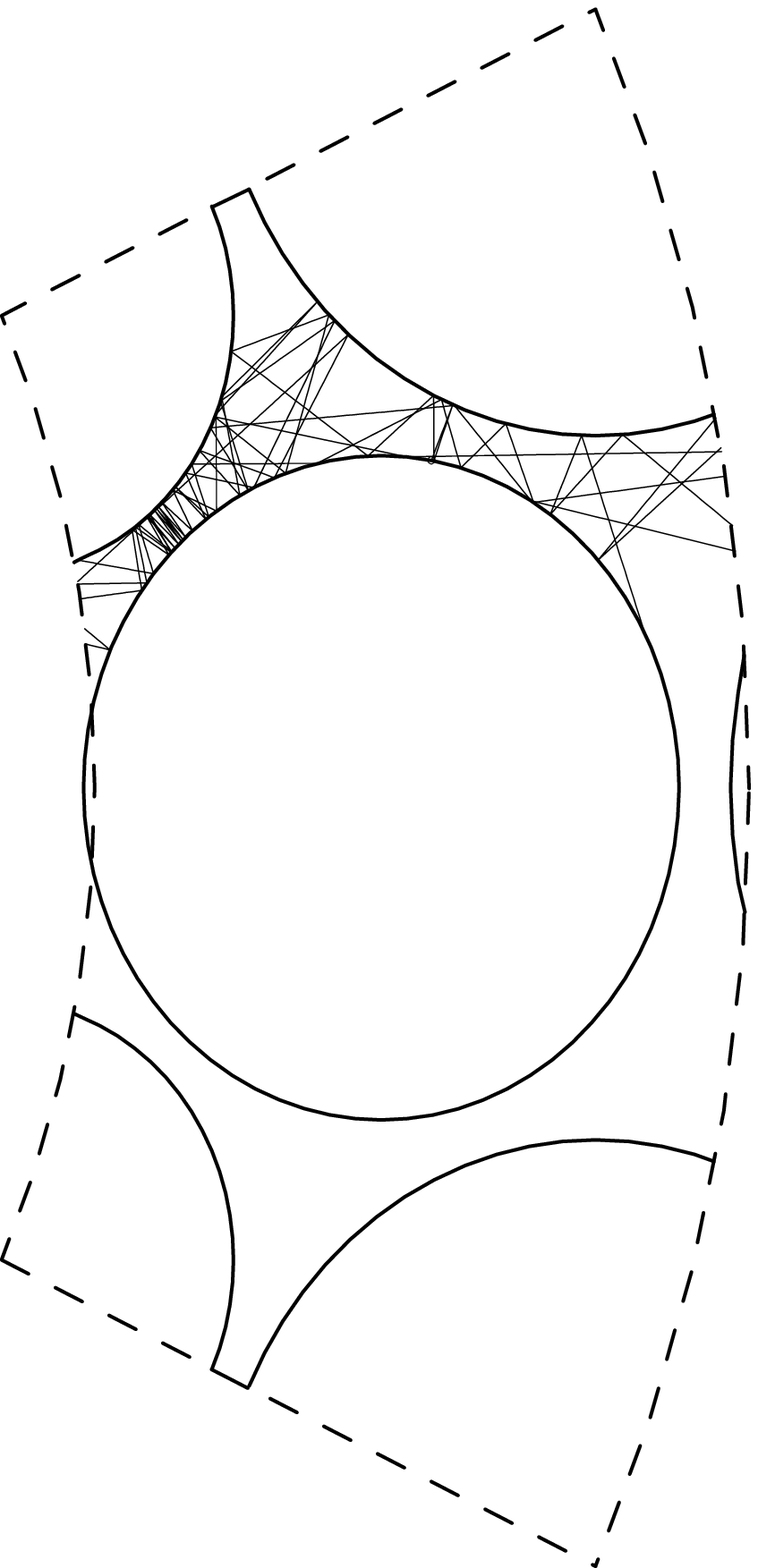}
\hspace{.1\textwidth}
\includegraphics[angle=0,width=.25\textwidth]{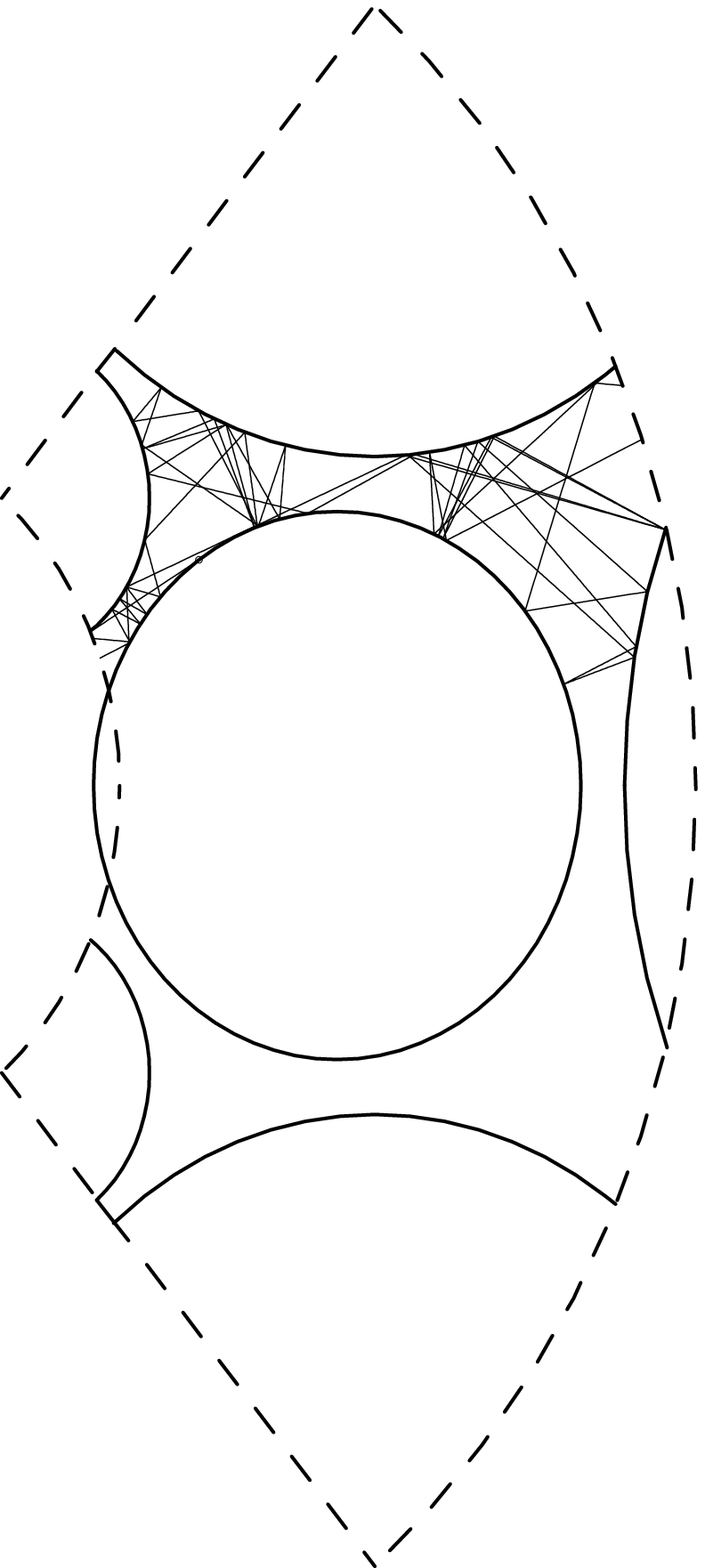}
\hspace{.1\textwidth}
\includegraphics[angle=0,width=.25\textwidth]{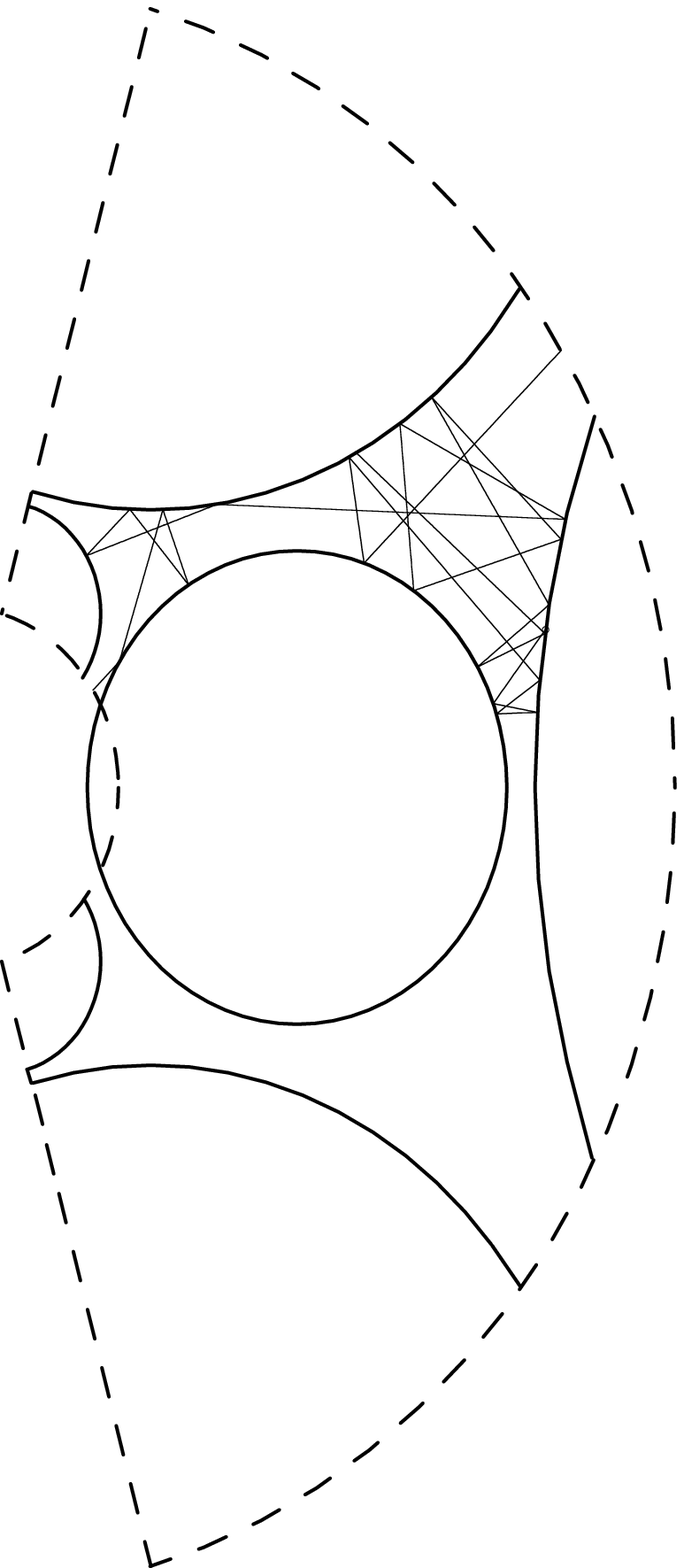}
\end{center}
\caption{Three examples of unit cells with short trajectories. The parameter
 values (defined in the text) are respectively $\epsilon = 0.5$ and $\rho
 = 0.459738$ (left),   $\epsilon = 1.$ and $\rho = 0.440862$ (center), and
 $\epsilon = 1.5$ and   $\rho = 0.413317$ (right).}
\label{fig.cell}
\end{figure}

In order to define the self-similar channel, we start with the description
of the reference cell, examples of which are shown in figure \ref{fig.cell}.

We define the reference cell as the interior of a region bounded on
the right and left sides by two arc-circles, with common center and
respective radii $\exp(\pm \epsilon/2)/\epsilon$, and, on the upper
and lower sides, by two oblique lines of slopes $\pm \tan \sqrt{3}
\epsilon/2$\footnote{We
 note that these slopes may well be vertical and beyond when $\epsilon \ge
 \pi / \sqrt{3}\simeq1.814$. This is not a restriction.}, with the
exclusion of obstacles, which are here taken to be regular
discs. The parameter $\epsilon$ will henceforth be referred to as the {\em
 scaling exponent}, in contrast to the parameter  $\mu$ which was defined
in \cite{BGR06,BG07a} (or $r$ in \cite{CD07}), and which we refer to as the
{\em scaling factor}, $\mu = \exp(\epsilon)$\footnote{The conformally
 transformed Gaussian iso-kinetic Lorentz channel has a similar cell
 geometry, but for the obstacles, which are deformed, slightly flattened,
 discs, see \cite{BG07b} for details. There the scaling exponent is given
 by the amplitude of the external forcing field.}.

Let the origin $(x,y) = (0,0)$ be on the middle of the arc-circle of the
left-hand side border. We have the central disc of radius $\rho$,
located at
$$(x,y) = \Big(\frac{1 - \exp(-\epsilon/2)}{\epsilon},0 \Big),$$
two discs of radii $\rho\exp(-\epsilon/2)$ on the inner circle, at
$$\Big(\frac{\exp(-\epsilon/2)}{\epsilon} [\cos(\sqrt{3}\epsilon/2) - 1],
\pm \frac{\exp(-\epsilon/2)}{\epsilon}\sin(\sqrt{3}\epsilon/2)\Big),$$
and two discs of radii $\rho\exp(\epsilon/2)$ on the outer circle, at
$$\Big(\frac{\exp(\epsilon/2)}{\epsilon} \cos(\sqrt{3}\epsilon/2) -
\frac{\exp(-\epsilon/2)}{\epsilon}, \pm
\frac{\exp(\epsilon/2)}{\epsilon}\sin(\sqrt{3}\epsilon/2)\Big).$$
Moreover, the corners where the boundaries of the cell intersect are
located exactly at the centers of the outer discs.

Let $\mathcal{D}_0$ denote the reference cell. The extended system is
obtained from the reference cell thus defined by
making a copy of it, denoted by $\mathcal{D}_1$, which we scale by a factor
of $\exp(\epsilon)$ and attach to the right-hand side of the reference
cell. Likewise, another copy of the reference cell, this time denoted by
$\mathcal{D}_{-1}$, is scaled by $\exp(-\epsilon)$ and attached to
its left-hand side. By repeating this procedure, copying and scaling the
right- and left-most cells and attaching these copies to the existing
sequence of cells, we obtain the self-similar Lorentz channel, which
consists of an infinite collection $\{\mathcal{D}_n\}_{n\in\mathbb{Z}}$ of
such cells. Obviously the usual Lorentz channel
\cite{Gas98} is recovered for $\epsilon = 0$.

Note that when the value of $\rho$ is large enough with respect to
$\epsilon$ , the central circle of the reference cell
$\mathcal{D}_0$ will intersect with its inner arc-circle border.
Correspondingly a sixth disc appears in the reference cell, whose
center is located in cell $\mathcal{D}_1$, to the right of the
reference cell (as is the case in the examples shown in figure
\ref{fig.cell}). This happens when $\rho \geq  [1 -
\exp(-\epsilon/2)]/\epsilon$.

A point-particle which moves in this extended system has unit speed and
bounces off elastically when it collides with the scatterers and upper and
lower walls. The circular walls which separate the cells from one another
have no incidence on the dynamics.

As noticed in \cite{BGR06}, the dynamics on the extended system can be
boiled down to a single cell. Periodic boundary conditions, with rescaling
of the arc-lengths and velocities, must then be imposed when trajectories
collide on the circular sides to the right and left, reappearing on the
opposite sides.

\section{Ergodic properties\label{sec.erg}}

Although the geometry of the self-similar Lorentz channel is similar
to that of the conformal transformation of the Gaussian iso-kinetic
Lorentz gas, their dynamics are very different. By taking circular
scatterers for all values of $\epsilon$, we ensure that no
transition to a non-hyperbolic regime will occur. Nevertheless the
similarities between the two lead to several immediate results
regarding the ergodic properties of self-similar billiards. We refer
the reader to \cite{CD07} for more details.

Two constraints have to be imposed on the parameter values, which have
straightforward expressions in this geometry.

The first constraint is the finite horizon condition, by which we avoid
the possibility that some trajectories be ballistic. It is a
straightforward generalization of the corresponding condition in the
periodic Lorentz gas and here becomes
\begin{equation}
\rho > \rho_\mathrm{min}\equiv\frac{\sin(\sqrt{3}\epsilon/4)}{\epsilon}.
\label{fh}
\end{equation}

The second condition is that two discs on the same radius (with
center at $\exp(-\epsilon/2)/\epsilon$) do not overlap so that the
cell remains connected,
\begin{equation}
\rho < \rho_\mathrm{max}\equiv
\frac{1}{\epsilon}\frac{\exp(\epsilon) - 1}{\exp(\epsilon) + 1}.
\label{nod}
\end{equation}

\begin{figure}[htb]
\begin{center}
\includegraphics[width=.6\textwidth]{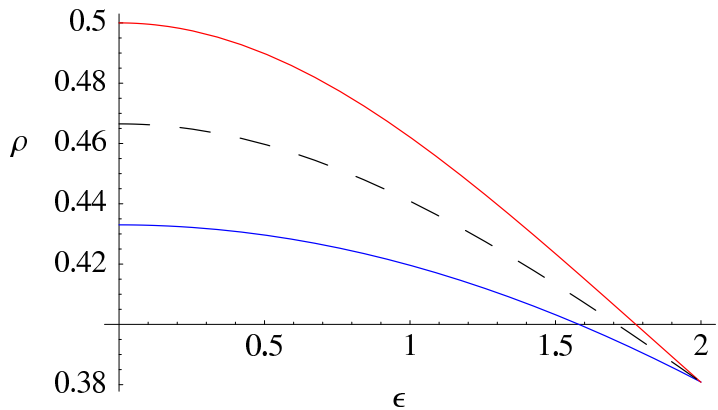}
\end{center}
\caption{Allowed values of $\rho$ v. $\epsilon$ from equations (\ref{fh})
 ($\rho$ above the blue curve), and (\ref{nod}) ($\rho$ below the red
 curve). Ergodicity will not be guaranteed outside the domain bounded by
 these two curves. The dashed line indicates the line of parameters we use
 in the computations presented in section \ref{sec.osc}.}
\label{fig.horizon}
\end{figure}

The parameter values compatible with these two conditions are shown in
figure \ref{fig.horizon}.

\section{Oscillations of the drift\label{sec.osc}}

Because of the biased geometry of the system, point-particles will
preferentially move from small cells to large cells, {\em i.~e.}\ from left
to right, thus inducing a current of mass along the horizontal axis
of the channel. Let $q(T)$ denote the displacement along this line
after a time $T$. On average, $q(T)$ grows linearly in $T$, which is
to say the current has a steady average value in time.

To be more precise, one can prove \cite{CD07} that the ratio
$q(T)/T$ remains of order one, in the sense that for any (small)
$p>0$ there are constants $0<a<b<\infty$ such that
$q(T)/T$ stays between $a$ and $b$ with probability $1-p$. However,
a detailed analysis \cite{CD07} shows the current may actually
retain a time dependence between those bounds.

The reason is that the probability distribution of the ratio
$q(T)/T$ does not stabilize as $T\to\infty$; rather it keeps
oscillating. This happens because the cell sizes grow exponentially;
thus the size of the current cell (that is the cell where the
particle is located at time $T$) is comparable to its entire
previous displacement $q(T)$. To emphasize this effect, assume for a
moment that the scaling factor between neighboring cells is $2$
({\em i.~e.} the scaling exponent $\epsilon=\ln 2$). Then the
current cell would be just as big as all the previous cells
combined. In other words, the geometric shape of the current cell
scales up at the same rate as the particle's entire path.

This observation should convince the reader that the geometric
structure of the current cell might affect the macroscopic evolution
of $q(T)/T$. For example, there may be some parts of the cell which,
on average, the particle traverses faster than others; in that case,
as it passes through `fast terrain', the ratio $q(T)/T$ would grow,
and as it drags itself through `slow motion areas', the ratio
$q(T)/T$ would decrease. This happens periodically, as the particle
moves from one cell to the next (even bigger) cell.

Note that the period of these oscillations is not fixed; it keeps
growing exponentially with $T$. More precisely, the length of the
period corresponds to the time it takes the particle to traverse the
current cell completely, thus the ratio $\log q(T) / \epsilon$ should
have period one.

These arguments were made precise in \cite{CD07}, where the following
theorem was proved~:

\begin{thm}
Assume $\epsilon$ is chosen such that equations (\ref{fh}) and (\ref{nod})
are satisfied. The following holds. Let $\{T_n\}_{n\in\mathbb{N}}$ be a
sequence of increasing times, $T_n\to\infty$, such that the fractional part
of $\log T_n/\epsilon$ tends to a constant $0\leq\delta<1$. Then the
distribution of $q(T_n)/T_n$ converges to a limit.
\label{thm.drift}
\end{thm}

The theorem also implies that the limit of $q(T_n)/T_n$ may depend
on the limiting fractional part of $\log T_n/\epsilon$, {\em i.~e.}\ of
$\delta$. We may therefore expect that $\langle q(T) \rangle/T$ will
display oscillations as a function of $\log T/\epsilon$, with unit
period. On the other hand, there may be no oscillations. Whether we
can observe this oscillatory regime or not depends on the parameter
values, as our numerical analysis reveals.

For the purpose of numerically demonstrating the drift oscillations, we let
the scaling exponent $\epsilon$ to vary between $0$ and its maximal
allowed value near $\epsilon = 2$ (see figure \ref{fig.horizon}) and take
the corresponding radius to be $\rho = 1/2(\rho_\mathrm{max} +
\rho_\mathrm{min})$, in between the two bounds specified by equations
(\ref{fh}) and (\ref{nod}). The precise values of the parameters we used in
our computations are shown in Table \ref{tab.param}.

\begin{table}
\caption{Parameter values used in the computations whose results are
 displayed in figures \ref{fig.drift}, \ref{fig.collapse} and
 \ref{fig.avgdrift}.}
\begin{tabular}{@{}ll|ll|ll|ll}
\br
$\epsilon$&$\rho$&$\epsilon$&$\rho$&$\epsilon$&$\rho$&$\epsilon$&$\rho$
\\
\mr
0.1&0.466231&0.6&0.456839&1.1&0.435921&1.6\lineup\0\0\0\0&0.407108\\
0.2&0.465406&0.7&0.453475&1.2&0.430665&1.7\lineup\0\0\0\0&0.400732\\
0.3&0.46404\lineup\0&0.8&0.44967\lineup\0&1.3&0.425125&
1.8\lineup\0\0\0\0&0.394213\\
0.4&0.462145&0.9&0.445456&1.4&0.419333&1.9\lineup\0\0\0\0&0.387575\\
0.5&0.459738&1.0&0.440862&1.5&0.413317&1.99765&0.380998\\
\br
\end{tabular}
\label{tab.param}
\end{table}

\begin{figure}[htb]
\begin{center}
\includegraphics[width=.3\textwidth]{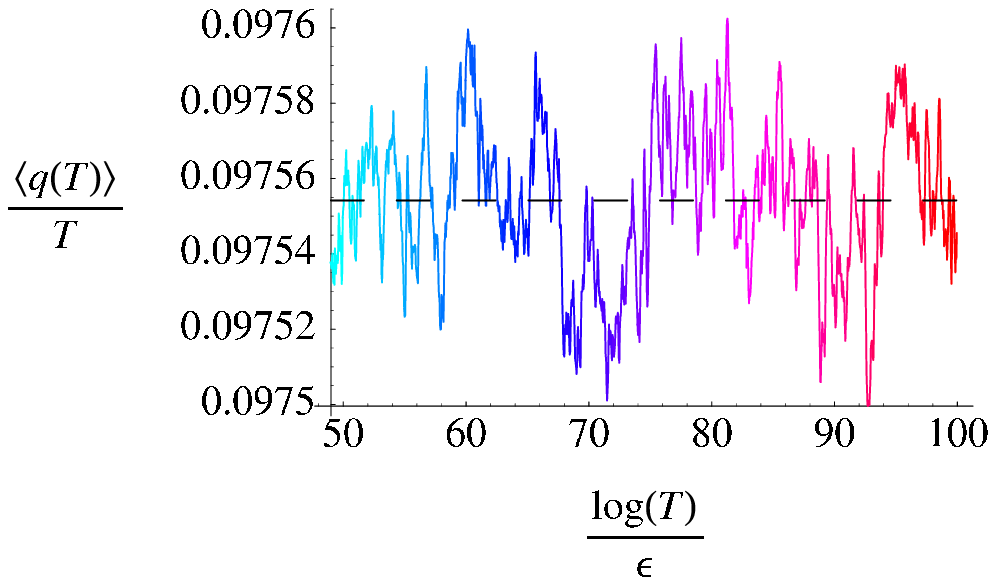}
\hspace{.025\textwidth}
\includegraphics[width=.3\textwidth]{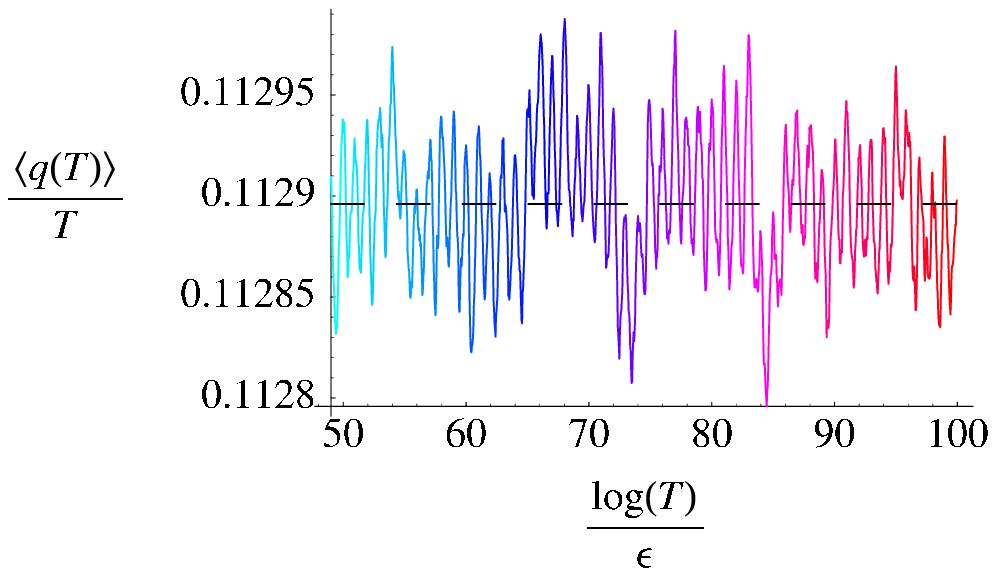}
\hspace{.025\textwidth}
\includegraphics[width=.3\textwidth]{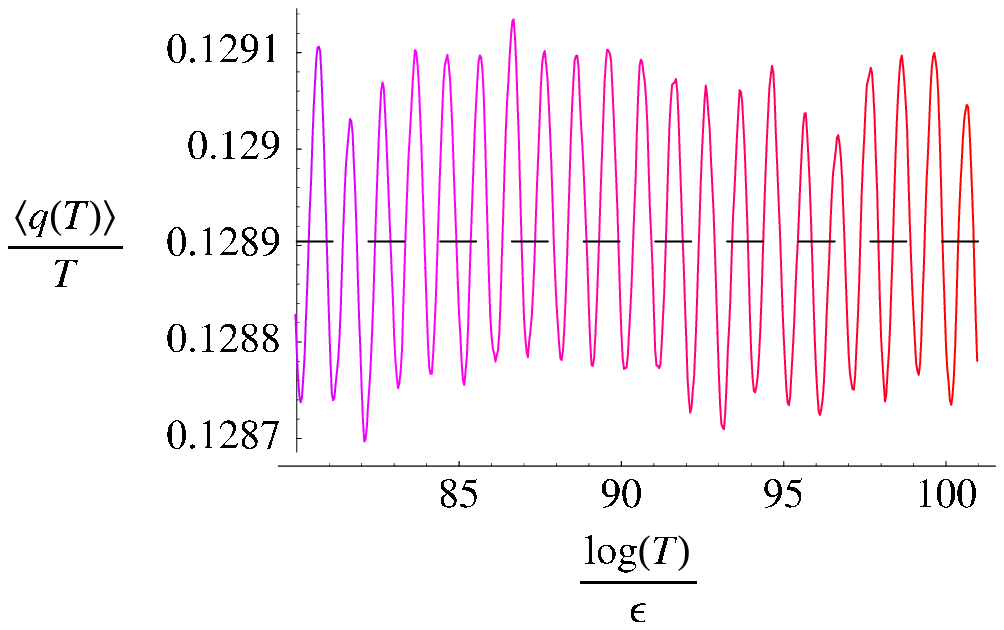}
\includegraphics[width=.3\textwidth]{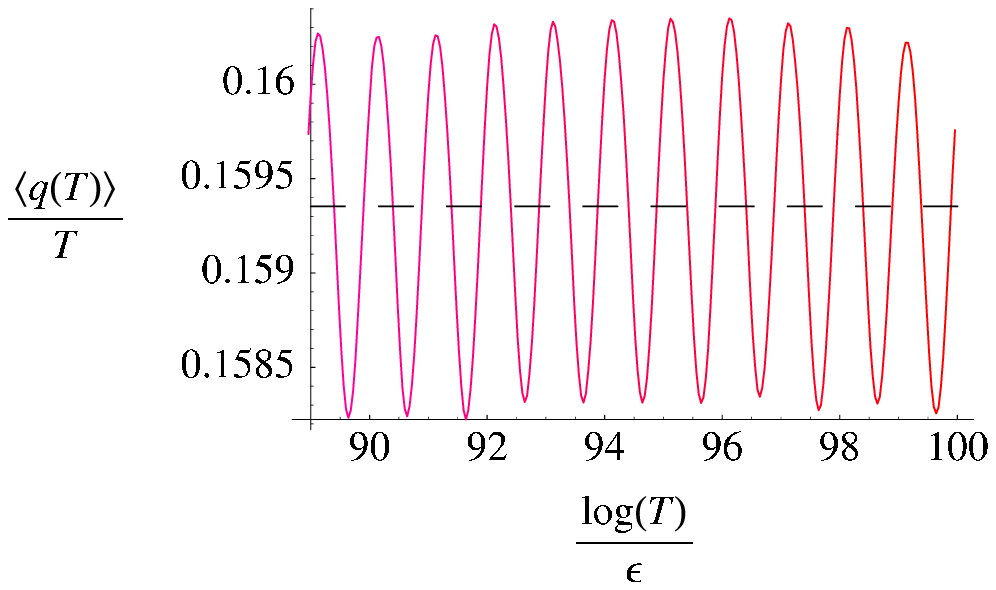}
\hspace{.025\textwidth}
\includegraphics[width=.3\textwidth]{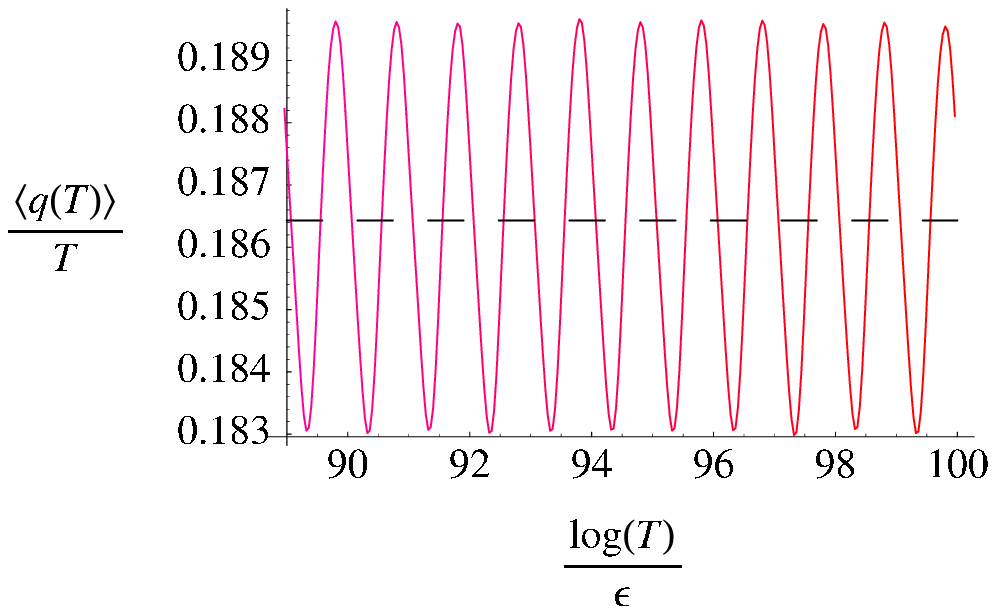}
\hspace{.025\textwidth}
\includegraphics[width=.3\textwidth]{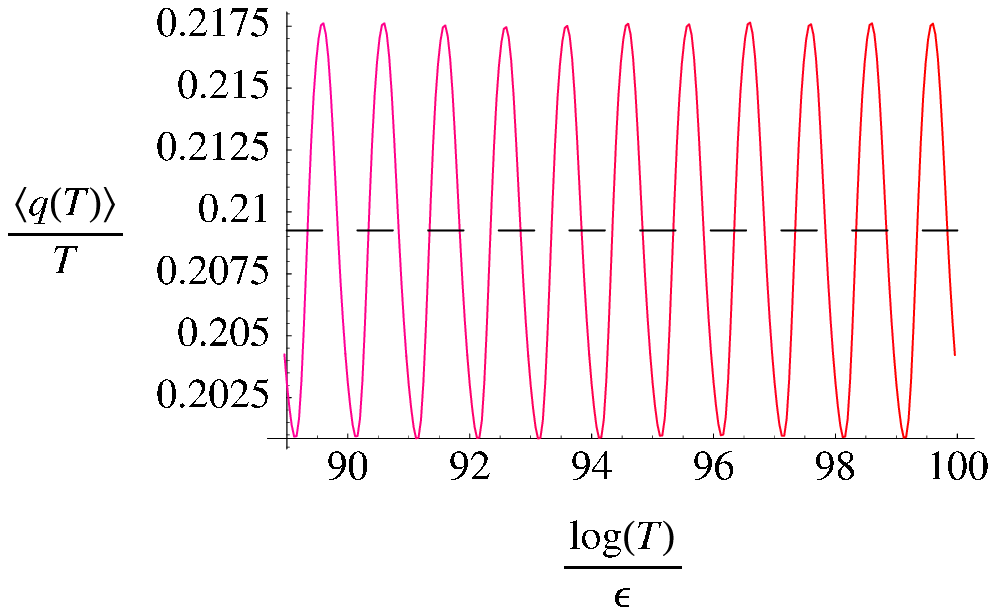}
\includegraphics[width=.3\textwidth]{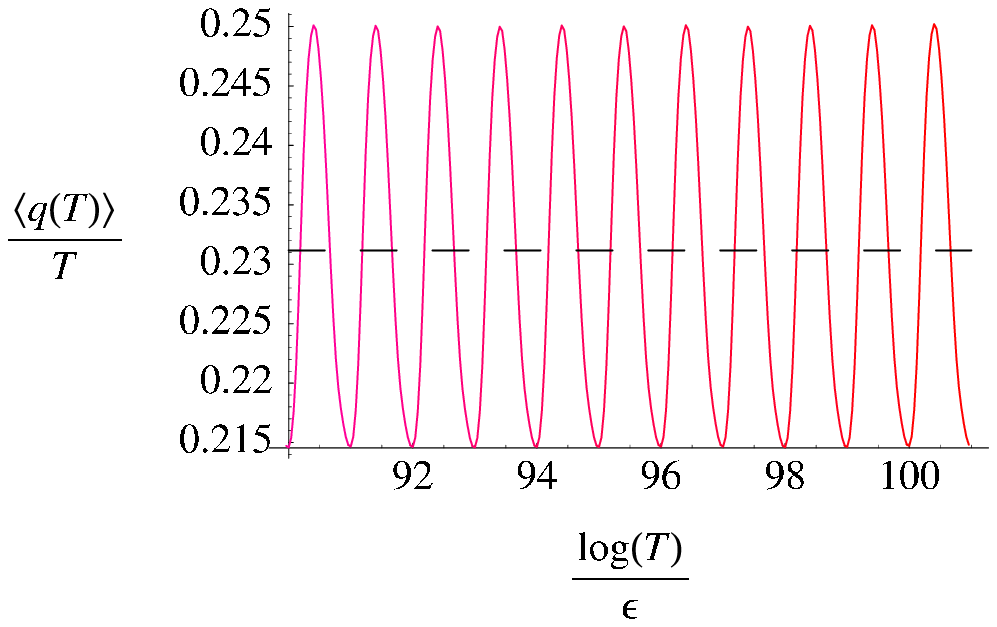}
\hspace{.025\textwidth}
\includegraphics[width=.3\textwidth]{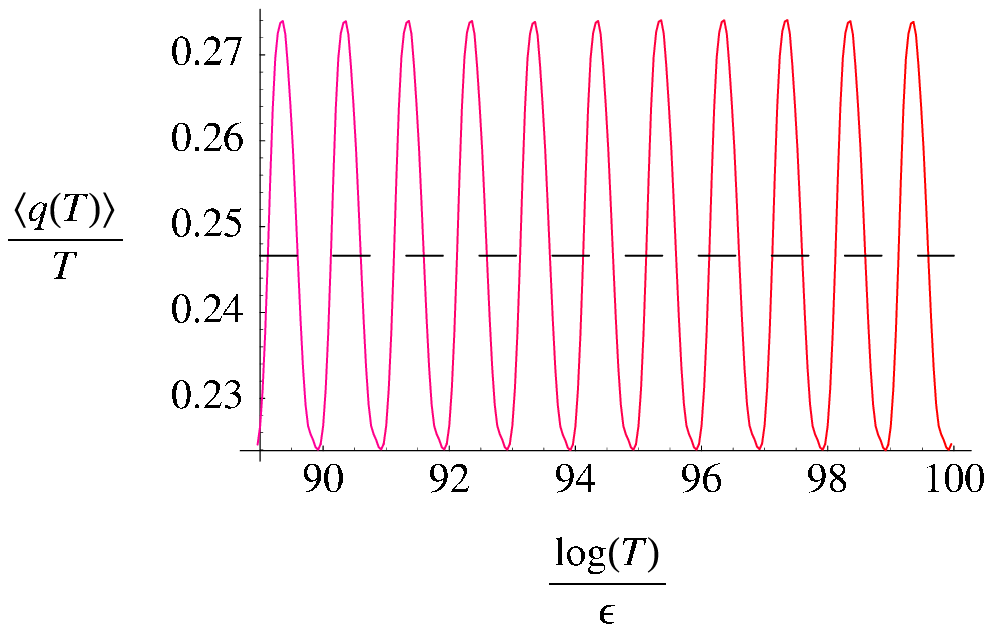}
\hspace{.025\textwidth}
\includegraphics[width=.3\textwidth]{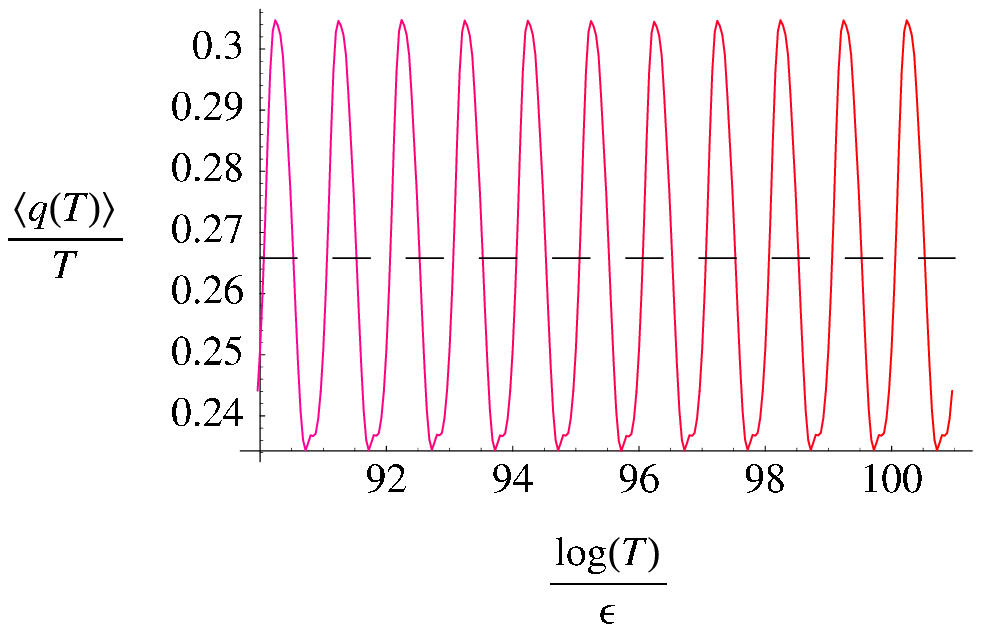}
\end{center}
\caption{Average drift terms $\langle q(T)\rangle/T$
 vs. $\log(T)/\epsilon$. Here shown are the parameter values of Table
 \ref{tab.param} corresponding to $\epsilon=0.8$, $0.9$, $1.0$, $1.2$,
 $1.4$, $1.6$, $1.8$, $1.9$ et $1.99765$, in order of increasing
 $\epsilon$, left to right and top to bottom. The dashed lines show the
 computed averages, as plotted in figure \ref{fig.avgdrift} below. The color
 code is a hue, regularly varying from cyan to red as units of $\log
 T/\epsilon$ increase from 50 to 100. Note that the span of the $x$-axis
 was reduced for the larger $\epsilon$ values in order to better display
 the periodicity of the oscillations.}
\label{fig.drift}
\end{figure}

Figure \ref{fig.drift} shows the results of numerical computations
of the drift function $\langle q(T)\rangle/T$ using $10^7$
trajectories with random initial conditions located on the central
circle (or on the discs on the horizontal line when there is an
overlap) of cell $\mathcal{D}_0$, and unit velocity at random angle.
For each of these trajectories, we computed the  horizontal position
$q(T)$ at times $T$ which we took to be logarithmically spaced on
the interval $0 \leq \log T/ \epsilon \leq 100$, so as to have $25$
points on each unit interval of the scale $\log T/\epsilon$.

\begin{figure}[htb]
\begin{center}
\includegraphics[width=.3\textwidth]{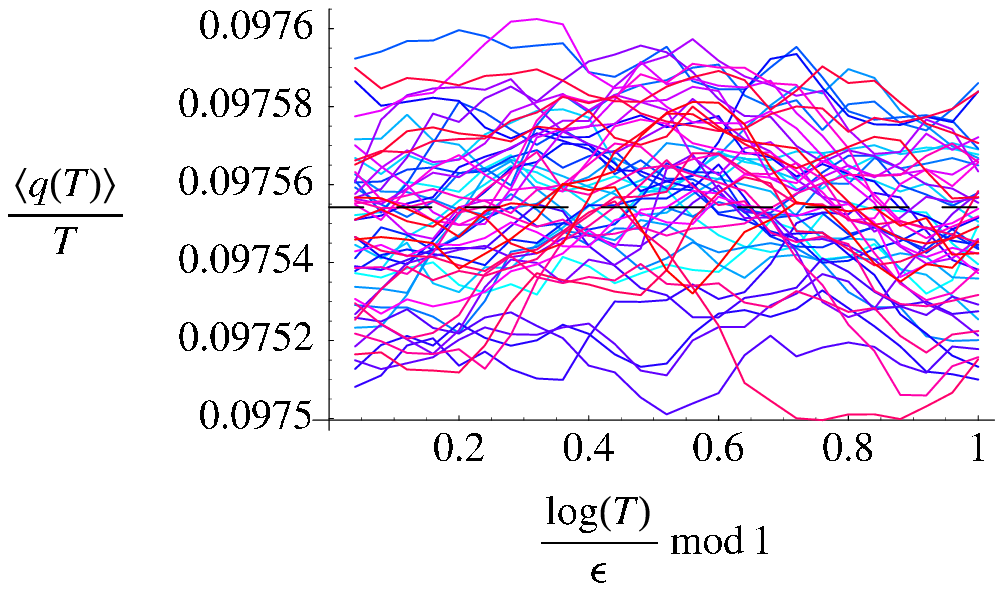}
\hspace{.025\textwidth}
\includegraphics[width=.3\textwidth]{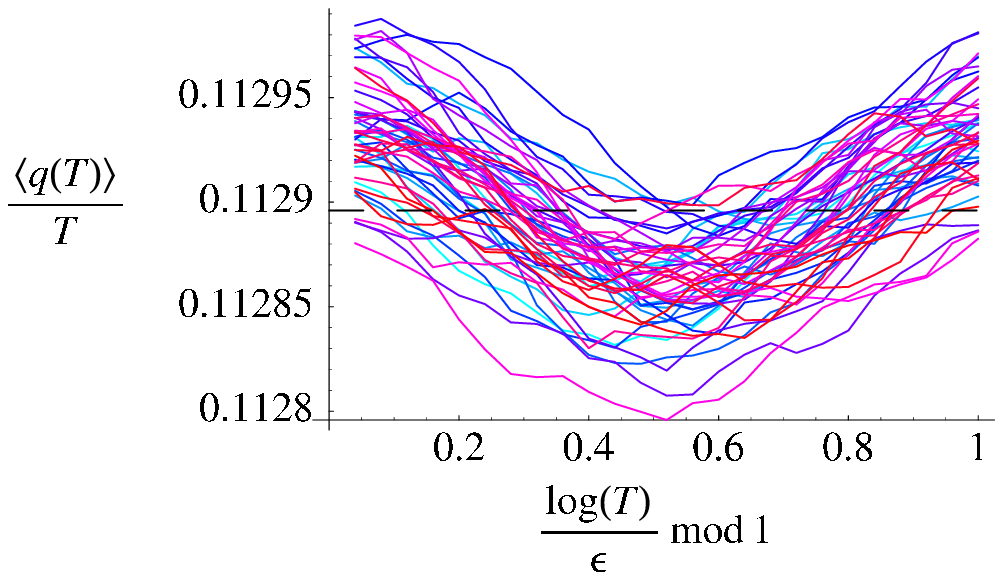}
\hspace{.025\textwidth}
\includegraphics[width=.3\textwidth]{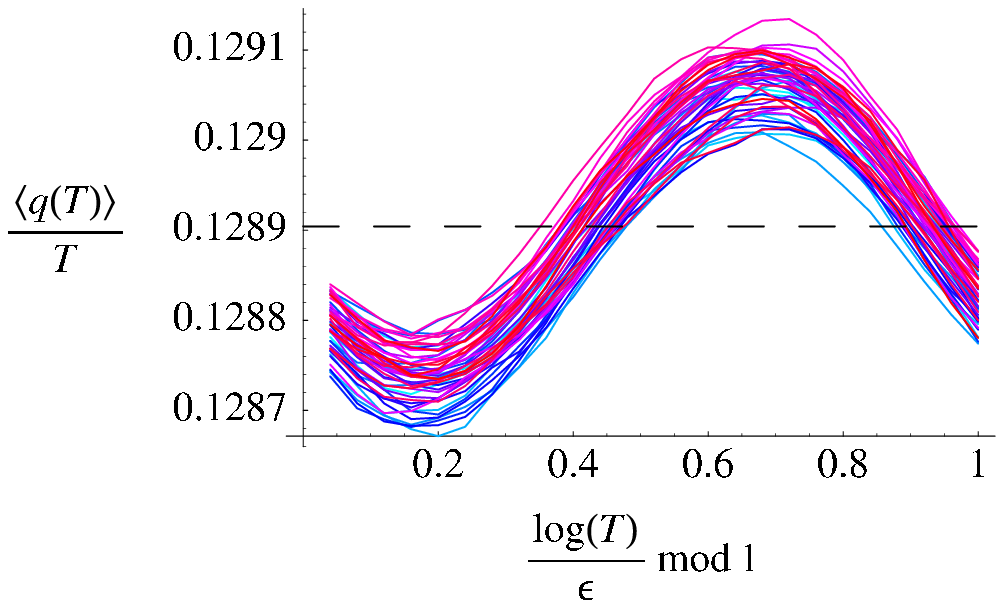}
\includegraphics[width=.3\textwidth]{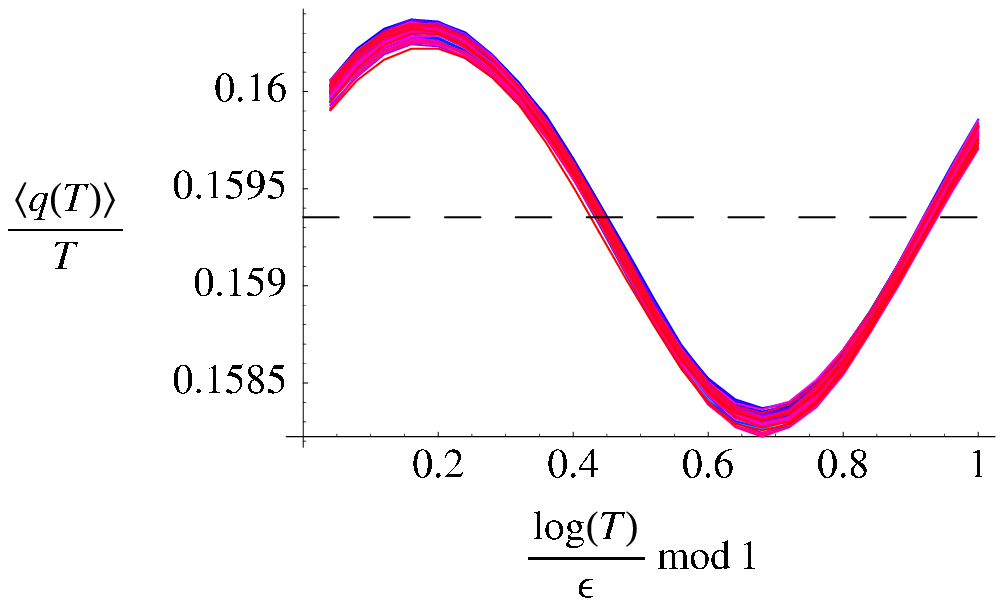}
\hspace{.025\textwidth}
\includegraphics[width=.3\textwidth]{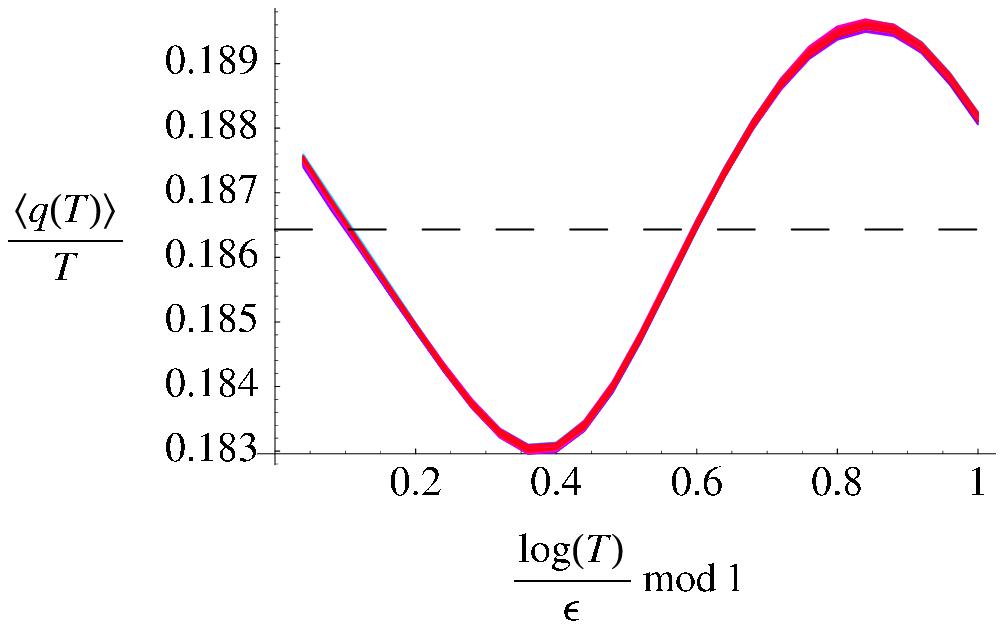}
\hspace{.025\textwidth}
\includegraphics[width=.3\textwidth]{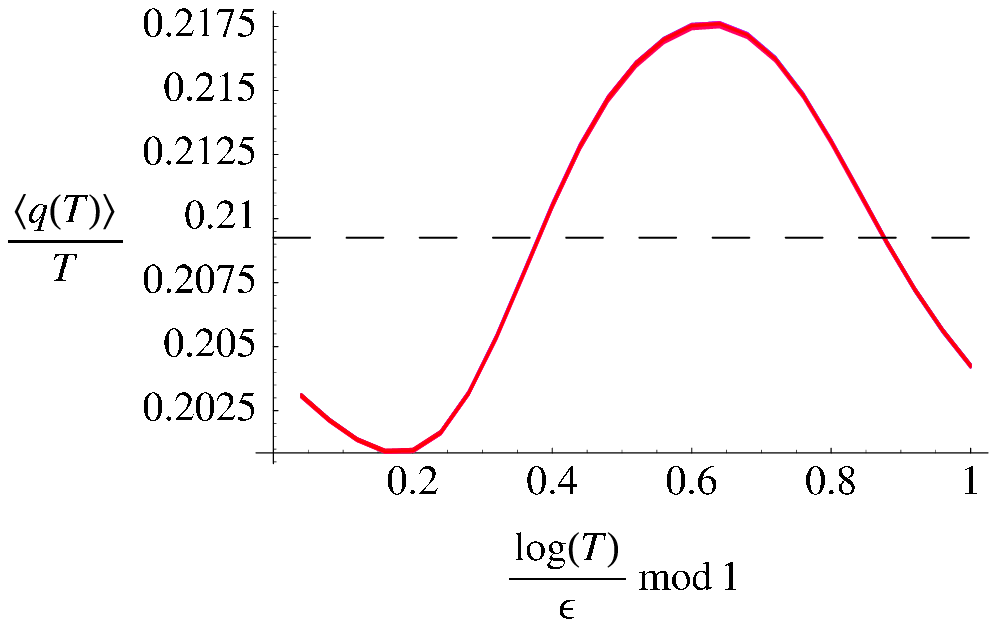}
\includegraphics[width=.3\textwidth]{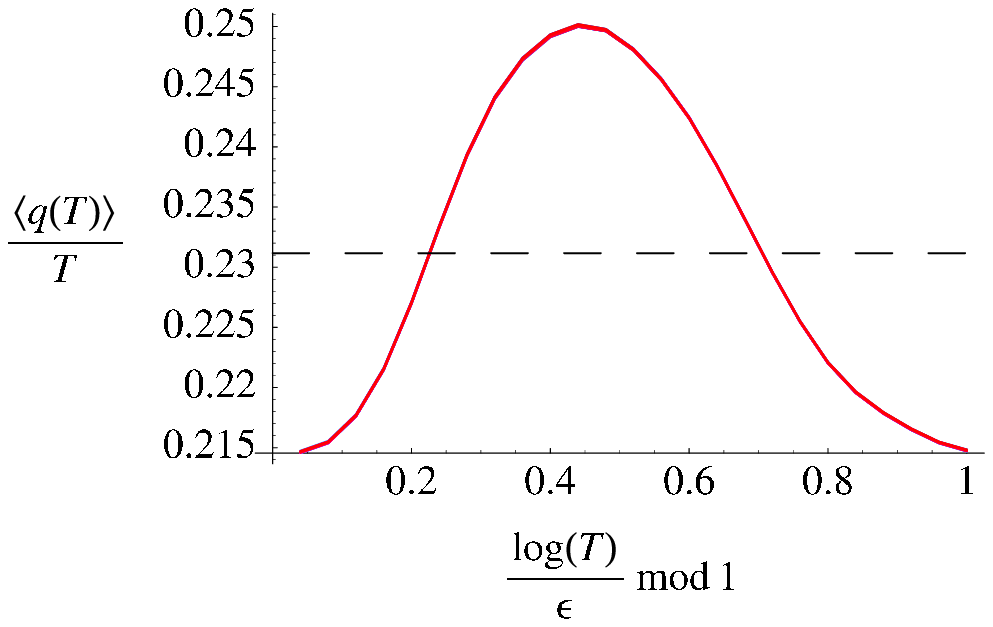}
\hspace{.025\textwidth}
\includegraphics[width=.3\textwidth]{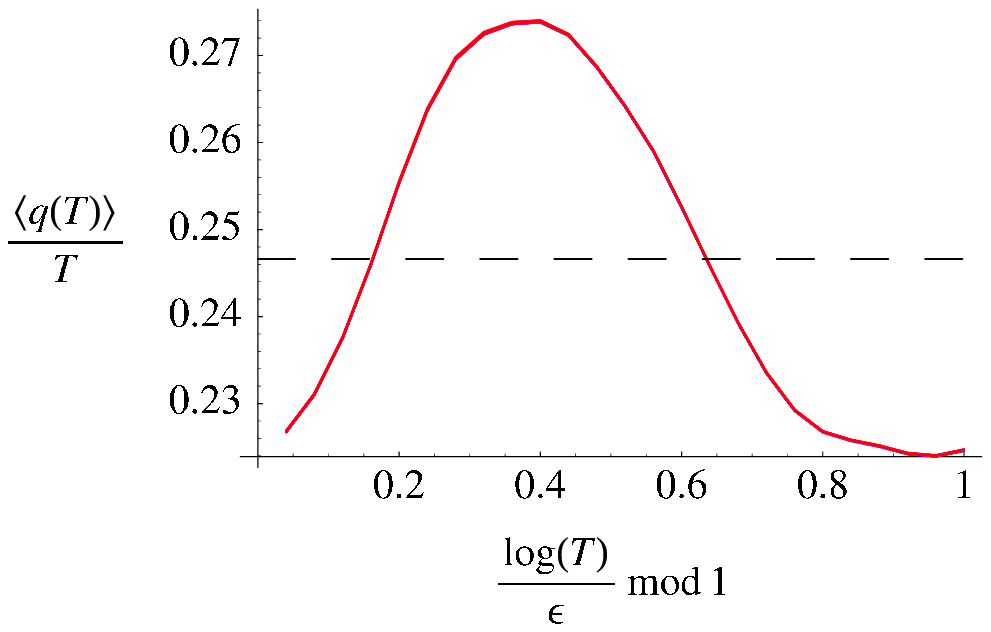}
\hspace{.025\textwidth}
\includegraphics[width=.3\textwidth]{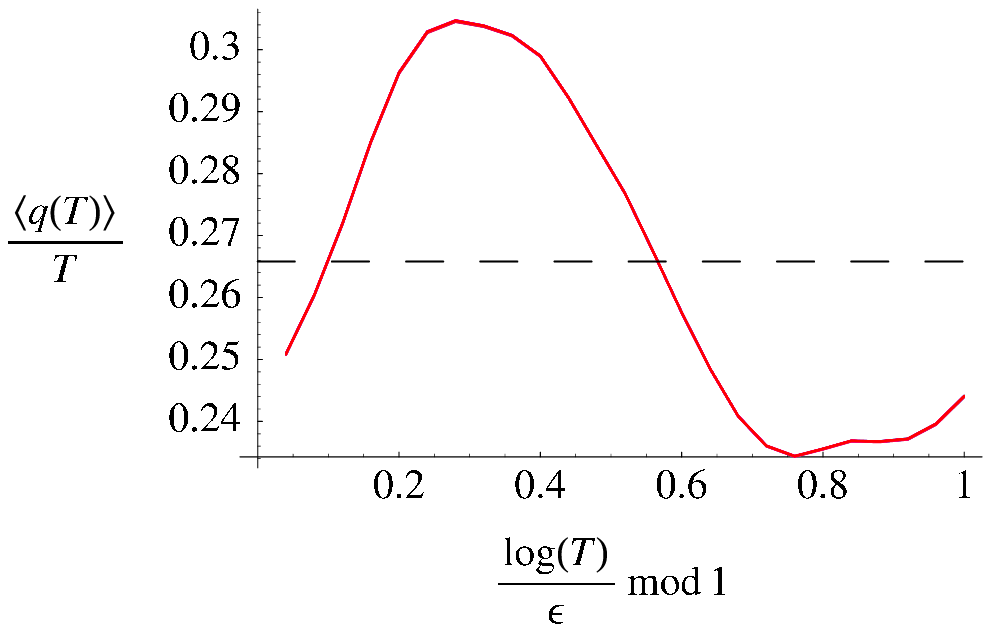}
\end{center}
\caption{Same as figure \ref{fig.drift} with the horizontal axis taken
 modulo 1. For every figure, we plot the last 50 oscillations. The color
 code is identical to figure \ref{fig.drift}. When $\epsilon$ is large
 enough, the curves nicely collapse on the same periodic function.}
\label{fig.collapse}
\end{figure}

For values of the scaling exponent $\epsilon \geq 1.0$, the
regularity of the observed periodic oscillations is spectacular, as
confirmed by the collapsed curves of figure \ref{fig.collapse}. On
the contrary, for values of $\epsilon \leq 0.8$ (only $\epsilon=0.8$
is shown on figures \ref{fig.drift} and \ref{fig.collapse}), our
measurements do not reveal periodic oscillations predicted by
theorem \ref{thm.drift}, which, in principle, applies to all values
of $\epsilon$. The lack of noticeable periodic oscillations for
small $\epsilon$ is, most likely, the effect of insufficient time
period $T$ in our simulation (recall that theorem \ref{thm.drift}
only holds asymptotically, as $T\to\infty$). It is (theoretically)
possible that, after many more (say, $10^3$ or $10^4$) initial
periods we would see proper periodic oscillations.

\begin{figure}[htb]
\begin{center}
\includegraphics[width=.45\textwidth]{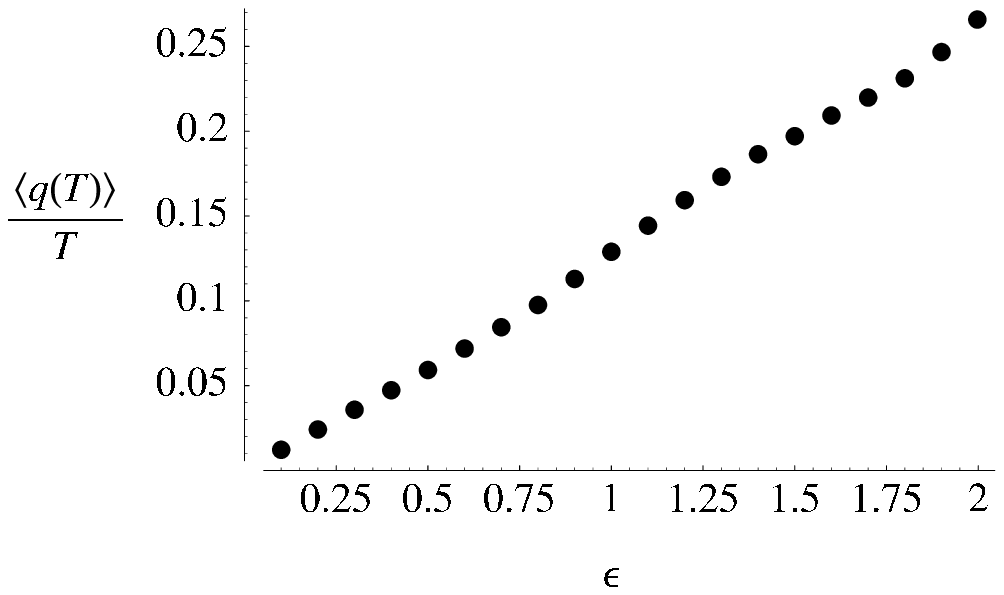}
\includegraphics[width=.45\textwidth]{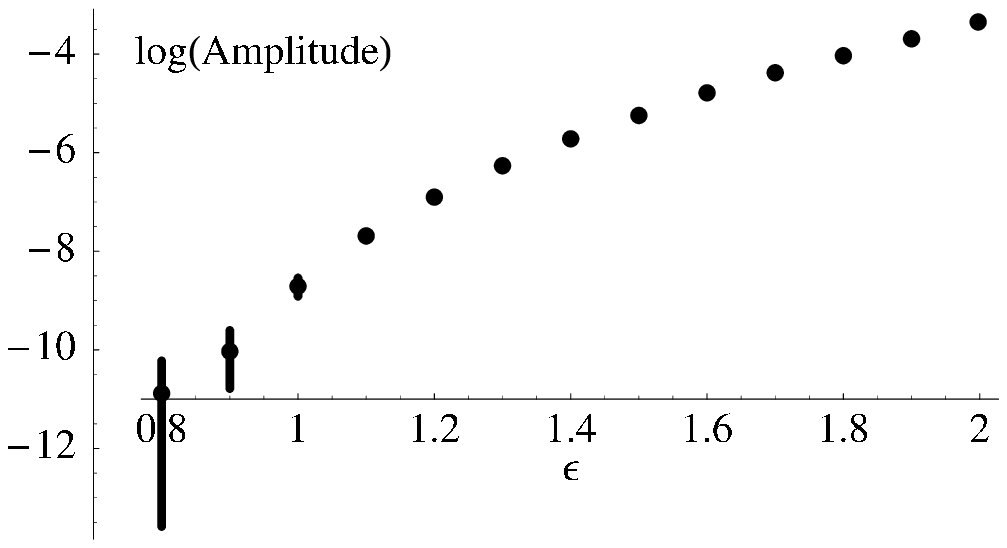}
\end{center}
\caption{Average drifts (left) and oscillation amplitudes (right)
 measured for the parameter values of table \ref{tab.param}.}
\label{fig.avgdrift}
\end{figure}

The summary of our computational results is shown in figure
\ref{fig.avgdrift}. The left-hand panel shows the average drift as a
function of $\epsilon$, computed by taking the average of the time
series of $\langle q(T)\rangle/T$ for every parameter value of Table
\ref{tab.param}. The right-hand panel shows the computed average
amplitude of the oscillations, which we denote by $A$, with the
corresponding error bars. The amplitudes of the oscillations is here
measured by taking the average of one half of the sum of the maximum
and minimum values of $\langle q(T)\rangle/T$ over each unit
interval of $\log T/\epsilon$, which corresponds to a period of
oscillation. Values of $\epsilon\leq0.7$ were discarded, because
oscillations could not be observed (yet).

The latter graph clearly demonstrates a power law behavior of the
amplitude in the scaling factor, {\em i.~e.}\ $A \sim \exp(a \epsilon)$,
with the coefficient $a\simeq 1.5$, for parameter values
$1.5<\epsilon<2.0$. The coefficient seems to be somewhat larger for
smaller values of $\epsilon$. The reason for this difference we
believe is due to a non-trivial dependence in the parameter $\rho$,
which here varies non-linearly with $\epsilon$. It might also be due to
our limited statistics (as also reflected by the large error bars).

\section{Concluding remarks \label{sec.con}}

To conclude, self-similar billiard channels are instances of
non-equilibrium chaotic billiards with volume-preserving dynamics,
whereby a geometric constraint induces a current of mass, going from
the smaller to the larger scales. The paper clearly establishes the
following three novel results~:
\begin{enumerate}
\item The average drift is a linear function in the scaling exponent $\epsilon$
 throughout its range of allowed values.
\item When the scaling exponent is large enough, log-periodic drift
 oscillations do occur, {\em i.~e.} the drift is not a constant.
\item The amplitude of the oscillations follows a power law in
 the scaling factor, more clearly so for the larger range of parameter
 values.
\end{enumerate}

It is to be noted that property (ii) is an expected consequence of
the discrete scale invariance of the system, which typically
manifests itself in the presence of power laws with complex
exponents. The signature of such power laws is the log-periodic
corrections to their scalings, in this case the corrections to
$q(T)\sim T$. Similar phenomena are observed in many different
physical situations, see \cite{Sor98} and references therein. That
the drift oscillations are barely noticeable when the scaling
exponent decreases below $\epsilon = 1.0$ is perhaps not surprising
as log-periodicity is typically a very small effect. Going to the
other end of the parameter range, it is actually remarkable that the
regularity of drift oscillations is so well pronounced for larger
values of $\epsilon$. In this regard, self-similar billiards offer
new ground to further studies of this interesting phenomenon, here
in the framework of hyperbolic dynamical systems, and thus improve
our understanding of a ubiquitous property which has already found
many known fields of applications.

\ack FB acknowledges financial support from Fondecyt project 1060820
and FONDAP 11980002 and Anillo ACT 15. NC is partially supported by
NSF grant DMS-0354775. TG  is financially supported by the Fonds
National de la Recherche Scientifique.


\section*{References}

\begin{thebibliography}{99}

\bibitem{H86} Hoover W G 2001 {\em Time Reversibility, Computer Simulation,
   and Chaos} (World Scientific, Singapore).

\bibitem{EM90} Evans D J and Morriss G P 1990 {\em Statistical Mechanics of
   Non-Equilibrium Fluids} (Academic Press, London).

\bibitem{Gas98} Gaspard P 1998 {\em Chaos, Scattering and Statistical
   Mechanics} (Cambridge University Press, Cambridge).

\bibitem{Dor99} Dorfman J R 1999 {\em An Introduction to Chaos in
   Nonequilibrium Statistical Mechanics} (Cambridge University Press,
 Cambridge).

\bibitem{CM06} Chernov N~I and Markarian R 2006 {\em Chaotic billiards}
 Math. Surveys and Monographs {\bf 127} (AMS, Providence, RI).

\bibitem{BSC91} Bunimovich L A, Sinai Ya G, and Chernov N I  1991
 {\em Statistical properties of two-dimensional hyperbolic billiards}
 Russ. Math. Surv. {\bf 46} 47.

\bibitem{GCGD01} Gaspard P, Claus I, Gilbert T, and Dorfman J R 2001
 {\em The fractality of the hydrodynamic modes of diffusion}
 Phys. Rev. Lett. {\bf 86} 1506.

\bibitem{Gas97} Gaspard P 1997 {\em Chaos and hydrodynamics} Physica A
 {\bf 240} 54.

\bibitem{CELS93} Chernov N I, Eyink G L, Lebowitz J L, and Sinai Ya G 1993
 {\em Derivation of Ohm's law in a deterministic mechanical model}
  Phys. Rev. Lett. {\bf 70} 2209; 1993
 {\em Steady state electric conductivity in the periodic Lorentz gas}
  Comm. Math. Phys. {\bf 154} 569.

\bibitem{R96} Ruelle D 1996
 {\em Positivity of entropy production in non-equilibrium statistical
   mechanics} J. Stat. Phys. {\bf 85} 1; 2003
 {\em Extending the definition of entropy to non-equilibrium steady
   states} Proc. Nat. Acad. Sci. {\bf 100} 30054.

\bibitem{DM96} Dettmann C~P and Morriss G~P 1996
 {\em Hamiltonian formulation of the Gaussian iso-kinetic thermostat}
 Phys. Rev E {\bf 54} 2495; Morriss G~P and Dettmann C~P 1998
 {\em Thermostats: analysis and application}
 Chaos {\bf 8} 321.

\bibitem{W00} Wojtkowski M~P 2000
 {\em W-flows on Weyl manifolds and Gaussian thermostats}
 J. Math. Pures Appl. {\bf 79} 953.

\bibitem{BG07b} Barra F and Gilbert T 2007 {\em Non-equilibrium Lorentz gas
   on a curved space} J. Stat. Mech. {\bf 7} L01003.

\bibitem{BGR06} Barra F, Gilbert T and Romo M 2006 {\em Drift of particles
   in self-similar systems and its Liouvillian interpretation}
 Phys. Rev. E {\bf 73} 026211.

\bibitem{BG07a} Barra F and Gilbert T 2007 {\em Steady-state conduction in
 self-similar billiards} Phys. Rev. Lett. {\bf 98} 130601.

\bibitem{CD07} Chernov N and Dolgopyat D 2007 {\em Particle's drift in
   self-similar billiards} preprint.

\bibitem{Sor98} Sornette D 1998 {\em Discrete scale invariance and complex
   dimensions} Phys. Rep. {\bf 297} 239.

\end{thebibliography}
\end{document}